\documentstyle[12pt]{article}
\def\da{{\dot{\alpha}}}
\def\db{{\dot{\beta}}}
 \def\a{\alpha}
\def\b{\beta}
\def\rar{\rightarrow}
\def\bchi{\bar{\chi}}
\def\bpsi{\bar{\psi}}
\def\sm{{\sigma_m}}
\def\sn{{\sigma_n}}
\def\sk{{\sigma_k}}

\def\dg{{\dagger}}
\def\bsm{{\bar{\sigma}_m}}
\def\bsn{{\bar{\sigma}_n}}

\def\bsl{{\bar{\sigma}_l}}
\def\e{\epsilon}
\def\le{\left(}
\def\ri{\right)}
\def\t{\theta}
\def\bt{{\bar{\theta}}}
\def\Rar{\Rightarrow}
\def\la{\lambda}
\def\bla{{\bar{\lambda}}}
\def\ve{\varepsilon}
\def\bve{{\bar{\varepsilon}}}
\def\Qb{{\bar{Q}}}
\def\pd{\partial}
\def\pdb{{\bar{\partial}}}
\def\ol{\overleftarrow}
\def\Db{{\bar{D}}}
\def\g{{\gamma}}
\def\yb{{\bar{y}}}
\def\smn{{\sigma_{mn}}}
\def\Dc{{\cal D}}
\def\f12{\frac{1}{2}}
\def\fra1g2{\frac{1}{g^2}}
\def\Tr{{\rm Tr}}

\def\Lac{\Lambda}
\def\bLac{{\bar{\Lambda}}}
\def\dis{\displaystyle}
\def\del{\delta}
\def\G{\Gamma}
\def\F{\Phi}
\def\bF{\bar{\Phi}}
\def\bW{\bar{W}}
\def\yt{\tilde{y}}  
\def\ytb{\bar{\tilde{y}}}
\def\Ab{\bar{A}}
\def\yb{\bar{y}}
\def\Mt{\widetilde{M}}
\def\Mtb{\overline{\widetilde{M}}} 
\def\Bb{\bar{B}}
\def\Mb{\bar{M}}
\def\gt{\tilde{g}}
\def\no{\nonumber}
\def\Vt{\tilde{V}}
\def\Kt{\tilde{K}}

\begin{document}
\begin{titlepage}
\flushright{SISSA/12/00/EP}

\vspace{2cm}
\begin{center}
{\Large \bf Renormalizations in softly broken N=1 theories: \\
\vspace{3mm} Slavnov--Taylor identities} 
\vglue 10mm
 Igor Kondrashuk\footnote{E-mail: ikond@sissa.it, 
on leave of absence from LNP, JINR, Dubna, Russia}
\vglue 5mm
{\it SISSA -- ISAS  and INFN, Sezione di Trieste, \\
 Via Beirut 2-4, I-34013, Trieste, Italy}
\end{center}
\vglue 20mm
\begin{abstract}
Slavnov--Taylor identities have been applied to perform explicitly the 
renormalization procedure for the softly broken N=1 SYM. The result is in 
accordance with the previous results obtained at the level of 
supergraph technique.   
\end{abstract}

\begin{center}
\end{center}

\end{titlepage}

\section{Introduction}

One of the ways to break supersymmetry is to introduce into the 
supersymmetric theory interactions with background superfields 
that are space-time independent. The relation between the 
theory with softly broken supersymmetry and 
its rigid counterpart has been studied in Refs. (\cite{Yam}-\cite{Giudice-2}).
The investigation has been performed for singular parts 
of the effective actions of softly broken and rigid theories. 
Since the only modification of the classical action from the rigid 
case to the softly broken case is a replacement of coupling constants 
of the rigid theory with background superfields,
the relation is simple and can be reduced to substitutions of  
these superfields into renormalization constants of the rigid theory 
instead of the rigid theory couplings \cite{Jones,AKK}.      
Later, a relation between full correlators of softly broken
and unbroken SUSY quantum mechanics has been found \cite{IKQM}.
More recently, nonperturbative results for the terms of the effective action 
which correspond to the case when chiral derivatives do not act on  
background superfields have been derived \cite{Ratt}. 

The renormalization of the soft theory has been made 
on the basis of supergraph technique in the Ref.\cite{AKK}. 
Here we perform the renormalization procedure for the softly broken theory 
using Slavnov--Taylor identities.    

The notation used for the D4 supersymmetry and for the classical 
action $S^{\rm{R}}$ (${\rm{R}}$ means ``rigid'')
of the theory without softly broken supersymmetry is given in the 
Appendix. To have a possibility to compare with the case of softly 
broken supersymmetry the renormalization procedure for the rigid $\rm N=1$
SYM is reviewed in the Appendix. 

\section{N=1 Softly Broken Theories}

The classical action $S^{\rm{S}}$ (the superscript ${\rm{S}}$ means ``soft'')
with softly broken supersymmetry repeats the rigid action $S^{\rm{R}}$ 
(\ref{SR}) except 
for the replacement couplings of the theory with background $x$-independent 
superfields,   
\begin{eqnarray}
& S^{\rm{S}} = \dis{\int d^4y d^2\t~S\frac{1}{2^7}\Tr~ W_\a W^\a  +  
\int d^4\yb d^2\bt~\bar{S}\frac{1}{2^7}\Tr~ \bW^\da \bW_\da } \nonumber \\
& + \dis{\int d^4x d^2\t d^2\bt ~ \bF^i{(e^{V})_i}^j{K_j}^k\F_k}  \label{SS}\\
& + \dis{\int d^4yd^2\t \left[\yt^{ijk}\F_i\F_j\F_k + \Mt^{ij}\F_i\F_j \right]} 
+ \dis{\int d^4\yb d^2\bt \left[\ytb_{ijk}\bF^i\bF^j\bF^k    
+ \Mtb_{ij}\bF^i\bF^j \right]}.  \no
\end{eqnarray}
The indices of the matter superfields are reducible. They run over  
irreducible representations and members of them. The external background 
$x$-independent superfields $S,$ $K_i^j,$ and $\dis{\yt_{ijk}}$ are   
\begin{eqnarray*}
& \dis{S = \fra1g2\le 1 - 2m_A\t^2 \ri},~~~
\dis{\bar{S} = \fra1g2\le 1 - 2\bar{m}_A\bt^2 \ri }, \\
& \dis{{K_i}^j = \del_i^j + \le m^2 \ri_i^j}\t^2\bt^2, \\
& \dis{\yt_{ijk} = y_{ijk} + A_{ijk}\t^2},~~~ 
 \dis{\ytb_{ijk} = \yb_{ijk} + \Ab_{ijk}\bt^2}, \\
& \dis{\Mt_{ij} = M_{ij} + B_{ij}\t^2},~~~ 
 \dis{\Mtb _{ij}= \Mb_{ij} + \Bb_{ij}\bt^2}.
\end{eqnarray*}

These superfields break supersymmetry in a soft way since they are 
not included in the supersymmetry transformation at the component level.  

\section{Slavnov--Taylor Identities}

In the rest of the paper we concentrate on the gauge part of the action. The 
renormalization of the chiral matter superfields is trivial and is evident from the 
supergraph technique \cite{Yam,AKK}.  

To fix the gauge we have to add the gauge fixing term and the ghost terms to the 
action (\ref{SS}) which we choose in a slightly different manner in comparison with 
the rigid case (\ref{gfgR}),  
\begin{eqnarray*}
& \dis{\int d^4x d^2\t d^2\bt ~\frac{1}{16}\Tr~\le\Db^2 \frac{V}{\sqrt{\tilde{\a}}}
\ri\le D^2 \frac{V}{\sqrt{\tilde{\a}}}\ri}
 \\ 
& + \dis{\int d^4y d^2\t~\frac{i}{2}\Tr~  b~\Db^2 
\le\frac{\del_{\bar{c},c}V}{\sqrt{\tilde{\a}}}\ri + 
\int d^4\yb d^2\bt~\frac{i}{2}\Tr~  \bar{b}~D^2
 \le\frac{\del_{\bar{c},c}V}{\sqrt{\tilde{\a}}}\ri.}
\end{eqnarray*}
where $b$ and $\bar{b}$ are antighost chiral and antichiral superfields, 
and $c$ and $\bar{c}$ are ghost chiral and antichiral superfields, respectively. 
Everywhere in this paper we consider the non-zero highest components of the couplings 
as an insertion into the rigid theory supergraphs. Such a choice of the gauge 
fixing term and the ghost terms means that we fix the gauge arbitrariness by 
imposing the condition 
\begin{eqnarray}
D^2\frac{V(x,\t,\bt)}{\sqrt{\tilde{\a}}}  = \bar{f}(\yb,\bt),  ~~~  
\Db^2 \frac{V(x,\t,\bt)}{\sqrt{\tilde{\a}}} = f(y,\t), \label{mod} 
\end{eqnarray}
where  $f$  and $\bar{f}$  are arbitrary chiral and antichiral functions.
This allows us to consider the gauge fixing constant $\tilde{\a}$ as an external 
$x$-independent background superfield on the same foot with the soft couplings and 
the soft masses of the softly broken action (\ref{SS}). This modification of 
the gauge fixing 
condition is important even at the level of supergraph technique 
\cite{AKK}. As it will be clear below this modification  
is the necessary way to remove divergences from the effective action of the 
softly broken theory using Slavnov--Taylor identities.  

Hence, the total gauge part of the classical action (\ref{SS}) is  
\begin{eqnarray}
& S^{\rm{S}}_{\rm{gauge}} = \dis{\int d^4y d^2\t~S\frac{1}{2^7}\Tr~ W_\a W^\a  +  
\int d^4\yb d^2\bt~\bar{S}\frac{1}{2^7}\Tr~ \bW^\da \bW_\da } \no \\
& + \dis{\int d^4x d^2\t d^2\bt ~\frac{1}{16}
\Tr~\le\Db^2 \frac{V}{\sqrt{\tilde{\a}}}\ri\le D^2\frac{V}{\sqrt{\tilde{\a}}}\ri}
  \label{SSgauge} \\ 
& + \dis{\int d^4y d^2\t~\frac{i}{2}\Tr~  b~\Db^2 
\le\frac{\del_{\bar{c},c}V}{\sqrt{\tilde{\a}}}\ri  + 
\int d^4\yb d^2\bt~\frac{i}{2}\Tr~  \bar{b}~D^2 
\le\frac{\del_{\bar{c},c}V}{\sqrt{\tilde{\a}}}\ri.} \no
\end{eqnarray}

The action (\ref{SSgauge}) is invariant under the same BRST symmetry as 
the rigid gauge action (\ref{SSgauge}) is except for the transformation of the 
antighost superfields which is a little different from that we have in the 
rigid case (\ref{BRSTr})
\begin{eqnarray}
& \dis{e^V \rar e^{i\bar{c}\ve} e^V e^{ic\ve}},& ~~~~  
\dis{\del b = \frac{1}{32}\le\Db^2D^2\frac{V}{\sqrt{\tilde{\a}}}\ri\ve} \no \\
& \dis{c \rar c + ic^2\ve},& ~~~~ \dis{\del \bar{b} = \frac{1}{32}
\le D^2\Db^2\frac{V}{\sqrt{\tilde{\a}}}\ri\ve}, \label{BRSTs} \\
& \dis{\bar{c} \rar \bar{c} - i\bar{c}^2\ve}, & ~~~~  \no
\end{eqnarray}
with a Hermitian Grassmannian parameter $\ve$, $\ve^\dg = \ve.$ 

The path integral describing the quantum soft theory is defined in 
the same way as the path integral (\ref{pathR}) of the rigid theory is      
defined, 
\begin{eqnarray}
& \dis{Z[J,\eta,\bar{\eta},\rho,\bar{\rho},K,L,\bar{L}] = 
\int dV~dc~d\bar{c}~db~d\bar{b}~\exp i}\left[\dis{S^{\rm{S}}_{\rm{gauge}}} 
\right.  \label{pathS}\\
& \left. + \dis{2~\Tr\le JV + i\eta c + i\bar{\eta}\bar{c} + i\rho b + 
i\bar{\rho}\bar{b}\ri + 
2~\Tr\le iK\del_{\bar{c},c}V + Lc^2 + \bar{L}\bar{c}^2 \ri}\right]. \no
\end{eqnarray}
The third term in the brackets is the BRST invariant since the external 
superfields $K$ and $L$ are BRST invariant by definition. All fields
in the path integral are in the adjoint representation of the gauge group.
For the sake of brevity we omit the symbol of integration in the 
terms with external sources, keeping in mind that it is the full superspace 
measure for vector superfields and the chiral measure for chiral superfields.

The ghost equation that is a reflection of invariance of the 
path integral (\ref{pathS}) under the change of variables  
\begin{eqnarray*}
b \rar b + \ve,    ~~~   \bar{b} \rar  \bar{b} + \bve 
\end{eqnarray*}
with an arbitrary chiral superfield $\ve$ must be modified in comparison 
with the ghost equation of the rigid theory (\ref{ghostE}) 
taking into account the modified BRST transformation of the antighost field
(\ref{BRSTs}). As the result, two ghost equations can be derived 
\begin{eqnarray*}
\bar{\rho} - i\frac{1}{4}D^2\frac{1}{\sqrt{\tilde{\a}}}\frac{\del W}{\del K} = 0,  ~~~
\rho - i\frac{1}{4}\Db^2\frac{1}{\sqrt{\tilde{\a}}}\frac{\del W}{\del K} = 0.
\end{eqnarray*}

The Legendre transformation (\ref{Legendre}) that
has been done in the Appendix for the rigid case can be repeated 
here without changes. Taking into account the relations (\ref{defphi}) 
and (\ref{GW}), the ghost equations can be represented as 
\begin{eqnarray}
\frac{\del \G}{\del\bar{b}}  - \frac{1}{4}D^2\frac{1}{\sqrt{\tilde{\a}}}
\frac{\del \G}{\del K} = 0,  ~~~
\frac{\del \G}{\del b} - \frac{1}{4}\Db^2\frac{1}{\sqrt{\tilde{\a}}}
\frac{\del \G}{\del K} = 0. \label{ghostEs}
\end{eqnarray}
  
If the change of fields (\ref{BRSTs}) in the path integral (\ref{pathS}) is made
we get the Slavnov--Taylor identity as the result of invariance of the 
integral (\ref{pathS}) under a change of variables. There is complete 
analogy with the rigid case (\ref{STr}) except for a little difference 
caused by the modified transformation of the antighost superfield in (\ref{BRSTs}).
The Slavnov--Taylor identities for the theory (\ref{pathS}) are
\begin{eqnarray}
& \Tr\left[\dis{\frac{\del \G}{\del V}\frac{\del \G}{\del K} 
 - i \frac{\del \G}{\del c}\frac{\del \G}{\del L}  
+ i \frac{\del \G}{\del \bar{c}}\frac{\del \G}{\del \bar{L}} 
- \frac{\del \G}{\del b}\le\frac{1}{32}\Db^2D^2\frac{V}{\sqrt{\tilde{\a}}}\ri} 
\right. \label{STs}\\ 
& \left. - \dis{\frac{\del \G}{\del \bar{b}}\le\frac{1}{32}D^2\Db^2\frac{V}
{\sqrt{\tilde{\a}}}\ri}\right]=0. \no
\end{eqnarray}

\section{Renormalizations of the Softly Broken SYM}

The identities (\ref{ghostEs}) and (\ref{STs}) allow us to remove all possible
divergences from the effective action $\G$ by rescaling superfields and 
couplings in the classical action (\ref{SSgauge}). Indeed, the identity
(\ref{ghostEs}) restricts the dependence of $\G$ on the antighost 
superfields and on the external source $K$ to an arbitrary dependence on their 
combination 
\begin{eqnarray*}
\le b + \bar{b}\ri\frac{1}{\sqrt{\tilde{\a}}}  + K.
\end{eqnarray*}
This means that the corresponding singular part of the effective action is 
\begin{eqnarray*}
\int  d^4x d^2\t d^2\bt~2i~\Tr\le (b + \bar{b})\frac{1}{\sqrt{\tilde{\a}}} + K\ri~
\tilde{A}(x,\t,\bt),
\end{eqnarray*}
where  $\tilde{A}(x,\t,\bt)$ is a combination of $c,\bar{c},V.$ By index counting 
arguments we know that the singular part repeats the structure of the classical action 
(\ref{SSgauge}) up to coefficients. Hence, $\tilde{A}(x,\t,\bt)$ starts from the 
$\tilde{z}_1(c + \bar{c}),$ since $\G$ is Hermitian. Here $\tilde{z}_1$ is a
constant that can be found by using the supergraph technique.

Now we can compare the renormalization constants  $\tilde{z}_1$ and $z_1.$ 
The constant $z_1$ is obtained from $\tilde{z}_1$ by putting all  
higher components of the soft couplings, of the soft masses, and of the 
gauge fixing coupling 
$\tilde{\a}$ in the action (\ref{SS}) equal to zero. In this case $z_1$ is a 
little different constant than that  
is appeared in the Appendix, since that rigid theory (\ref{pathR}) has another 
gauge fixing condition. Taking into account arguments based on the 
index of divergence and keeping in mind the absence of chiral derivatives 
in the ghost parts of the actions (\ref{SRgauge}) and  (\ref{SSgauge})   
we can see that 
\begin{eqnarray}
& \dis{\tilde{z}_1\le \gt^2,\sqrt{\tilde{\a}}\ri = 
z_1\le g^2 \rar \gt^2, \sqrt{\a} \rar \sqrt{\tilde{\a}} \ri},  \label{rule}\\
& \dis{\gt^2 =  g^2 \le 1+ m_A\t^2 + \bar{m}_A\bt^2 + 2m_A\bar{m}_A\t^2\bt^2 \ri = 
\le\frac{S + \bar{S}}{2}\ri ^{-1}.} \no
\end{eqnarray}
The substitution $g^2 \rar \gt^2$ becomes obvious if we remember that 
we consider higher components of the gauge coupling as insertions 
into the vector propagator and into the vector vertices 
in supergraphs \cite{Yam,AKK}. In short words,
the arguments of Refs. \cite{Yam,AKK} are the following. Since the action of a 
chiral derivative on spurions means decreasing the index of divergence 
inherited from a rigid diagram, a supergraph with logarithmic divergence
becomes convergent in this case. Hence, for the divergent part 
all spurions must be taken out of a supergraph together with rigid couplings.

By the same reason we take out of a supergraph the external superfield 
$\sqrt{\tilde{\a}}.$ Under the condition 
\begin{eqnarray*} 
\tilde{\a} = \gt^2
\end{eqnarray*}
we get the result obtained in the Ref. \cite{AKK} at the supergraph level
for the renormalization constants that become $x$-independent vector 
superfields,      
\begin{eqnarray*} 
\tilde{z}_1 = z_1\le g^2 \rar \gt^2 \ri.
\end{eqnarray*}
 
In the same way as it takes place in the rigid case, the Slavnov--Taylor identity
(\ref{STs}) fixes the coefficient before the longitudinal part of the 2-point vector 
Green's function. Indeed, by using projectors from (\ref{proj}) the infinite 
part of the 2-point vector correlator can be decomposed as  
\begin{eqnarray}
& \dis{V\le D,\tilde{z}_{a},\Db,\tilde{z}_{b},D,\tilde{z}_{c},\Db,\tilde{z}_{d} \ri V = 
V\le D,\tilde{z}_{a},\Db,\tilde{z}_{b},D,\tilde{z}_{c},\Db,\tilde{z}_{d}
 \ri \frac{D^\a\Db^2D_\a}{8\Box} V} 
\label{decS} \\  
& - ~\dis{V \le D,\tilde{z}_{a},\Db,\tilde{z}_{b},D,\tilde{z}_{c},\Db,\tilde{z}_{d}
\ri\frac{D^2\Db^2 + \Db^2D^2}{16\Box}V}, \no
\end{eqnarray}     
where the four derivatives in parenthesis can stand in some (in general, unknown) 
way. The difference from the rigid case decomposition (\ref{decR}) of the 2-point 
vector correlator is in a possible presence of $x$-independent background superfields   
$\tilde{z}_{a},\tilde{z}_{b},\tilde{z}_{c},\tilde{z}_{d}$ between these derivatives. 

The identity (\ref{STs}) means that these four derivatives in the second term 
of this decomposition must cancel $\Box$ in the denominator and  
the longitudinal term is reduced to the form  
\begin{eqnarray*}
\tilde{z}_2\frac{1}{32} \frac{V}{\sqrt{\tilde{\a}}}\le D^2\Db^2 + \Db^2D^2\ri  
\tilde{z}_2\frac{V}{\sqrt{\tilde{\a}}}.   
\end{eqnarray*}
It is not difficult to check that the Slavnov--Taylor identity also gives 
that $\tilde{z}_2 = 1,$ that is, there is no infinite correction to the longitudinal 
part of the 2-point vector Green's function in the soft case. The same arguments can be 
applied even in the case of the total effective action, taking into account the whole 
dependence of the effective action $\G$ on the combination 
\begin{eqnarray*}
\le b +\bar{b}\ri\frac{1}{\sqrt{\tilde{\a}}}  + K. 
\end{eqnarray*}
Hence, there is no finite correction to the longitudinal part of the 2-point vector 
correlator in the soft case. 

Now it is necessary to consider contributions in $\tilde{A}(x,\t,\bt)$ of the next 
orders in fields. For example, the third order terms can be presented as 
\begin{eqnarray}
& \dis{\int  d^4x d^2\t d^2\bt~2i~\Tr\le \le b +\bar{b}\ri\frac{1}{\sqrt{\tilde{\a}}}
 + K\ri~\left[
\tilde{z}_1(c + \bar{c}) +\tilde{z}_4 \le Vc + \bar{c}V \ri 
+ \tilde{z}_5\le cV + V\bar{c}\ri \right]} \no \\
& + \dis{\int d^4y d^2\t~2\Tr~\tilde{z}_6Lc^2 + 
\int d^4\yb d^2\bt~2\Tr~\bar{\tilde{z}}_6\bar{L}\bar{c}^2} \label{3odS}
\end{eqnarray}
By the no-renormalization theorem for the superpotential \cite{West} we get 
\begin{eqnarray*}
\tilde{z}_6 = \bar{\tilde{z}}_6 = 1.
\end{eqnarray*}
To fix the constants $\tilde{z}_4$ and $\tilde{z}_5$, we make the change of 
variables in the effective action $\G$
\begin{eqnarray}
& \dis{\G\left[V,c,\bar{c},b,\bar{b},K,L,\bar{L}\right] = 
\G\left[V(\Vt),c,\bar{c},b,\bar{b},K(\Kt),L,\bar{L}\right] = 
\tilde{\G}\left[\Vt,c,\bar{c},b,\bar{b},\Kt,L,\bar{L}\right]}, \no\\
& \dis{V = \Vt \tilde{z}_1,  ~~~  K = \frac{\Kt}{\tilde{z}_1}.} \label{changeS}
\end{eqnarray}
The Slavnov--Taylor identity (\ref{STs}) in the new variables is 
\begin{eqnarray}
& \Tr\left[\dis{\frac{\del \tilde{\G}}{\del \Vt}\frac{\del \tilde{\G}}{\del \Kt} 
 - i \frac{\del \tilde{\G}}{\del c}\frac{\del \tilde{\G}}{\del L}  
+ i \frac{\del \tilde{\G}}{\del \bar{c}}\frac{\del\tilde{\G}}{\del \bar{L}} 
- \frac{\del\tilde{\G}}{\del b}\le\frac{1}{32}\Db^2D^2\frac{\Vt\tilde{z}_1}
{\sqrt{\tilde{\a}}}\ri} \right. 
\label{STsm}\\ 
& \left. - \dis{\frac{\del \tilde{\G}}{\del \bar{b}}
\le\frac{1}{32}D^2\Db^2\frac{\Vt\tilde{z}_1}{\sqrt{\tilde{\a}}}\ri}
\right]=0. \no
\end{eqnarray}
The part of the effective action (\ref{3odS}) in the new variables looks like 
\begin{eqnarray}
& \dis{\int  d^4x d^2\t d^2\bt~2i~\Tr\le (b + \bar{b})
\frac{\tilde{z}_1}{\sqrt{\tilde{\a}}} + \Kt\ri~\left[
(c + \bar{c}) + \tilde{z}_4' \le \Vt c + \bar{c}\Vt \ri 
+ \tilde{z}_5'\le c\Vt + \Vt\bar{c}\ri \right]} 
\no  \\
&  \dis{+~~\int d^4y d^2\t~2\Tr~Lc^2 + 
\int d^4\yb d^2\bt~2\Tr~\bar{L}\bar{c}^2}, \label{3odSm}
\end{eqnarray}
where $\tilde{z}_4'$ and $\tilde{z}_5'$ are new constants. 

The higher order terms in the brackets of (\ref{3odSm}) are restored 
unambiguously by themselves in the iterative way due to the first three terms 
in the modified identities (\ref{STsm}). As the result, we have 
\begin{eqnarray}
\dis{\int  d^4x d^2\t d^2\bt~2i~\Tr\le (b + \bar{b})
\frac{\tilde{z}_1}{\sqrt{\tilde{\a}}} + \Kt\ri~\left[
\del_{\bar{c},c}\Vt\right].}  \label{depS}
\end{eqnarray} 

Now it is necessary to consider the transversal part of the 2-point 
vector correlator. Having made the change of variables in the effective 
action (\ref{changeS}), we see that the only structures of derivatives 
in  the  2-point vector Green's function    
\begin{eqnarray*}
\dis{\int d^4x d^2\t d^2\bt~
 \tilde{z}_1\Vt\le D,\tilde{z}_{a},\Db,\tilde{z}_{b},D,\tilde{z}_{c},\Db,\tilde{z}_{d}
\ri \tilde{z}_1\Vt } 
\end{eqnarray*}
which are allowed by the modified identities (\ref{STsm}) are 
\begin{eqnarray}
& \dis{~\int d^4x d^2\t d^2\bt~S
\frac{1}{2^5}~f(S)~\le D_\a\Vt\ri\le\Db^2D^\a\Vt\ri + {\rm H.c.}} \label{2pointS} \\
& + \dis{\int  d^4x d^2\t d^2\bt~\Tr\frac{1}{32}\frac{\tilde{z}_1\Vt}
{\sqrt{\tilde{\a}}}\le D^2\Db^2 +
 \Db^2D^2\ri \frac{\tilde{z}_1\Vt}{\sqrt{\tilde{\a}}}}. \no 
\end{eqnarray}
Here we have used the dependence of the singular part of $\tilde{\G}$ 
on the external source $\Kt$ which has already been fixed by (\ref{depS}). 
The function $f$ must be a chiral superfield. 

Since the function $f$ is obtained from the background superfields in the case
when chiral derivatives do not act on them, it can be obtained as the 
result of the change of rigid theory couplings with 
background superfields. But we have only one chiral 
background superfield which is the soft gauge coupling $S.$  
Hence, $f(S)$ can be obtained from the corresponding coefficient
of the rigid theory by the change 
\begin{eqnarray*}
\fra1g2 \rar S. 
\end{eqnarray*}
In the limit of constant gauge coupling we have 
\begin{eqnarray*}
\dis{~\int d^4y d^2\t~\fra1g2
\frac{1}{2^7}~z_1^2~z_3~\le 
\Db^2D_\a\Vt\ri \le\Db^2D^\a\Vt\ri + {\rm H.c.},}
\end{eqnarray*}
where $z_1$ and $z_3$ are renormalization constants of 
the rigid theory. Hence, we can derive that 
\begin{eqnarray}
\dis{f(S)|_{\t^2 = 0}~ = z_3~z_1^2 = z_{g^2},  ~~~  
f(S) \equiv \tilde{z}_S(S) =  z_{g^2}\le\fra1g2 \rar S\ri.}    \label{rule2}
\end{eqnarray}    
Hence, the renormalization constants  $\le\tilde{z}_S,~z_{g^2}\ri$ are not 
related like in the rule (\ref{rule}) for the pair 
$\le\tilde{z}_1,~z_1\ri$, but are related in the holomorphic way (\ref{rule2}).

The first term in the modified identity (\ref{STsm}) will restore in the 
iterative way higher order terms starting from the bilinear transversal 
2-point correlator (\ref{2pointS}). Hence, the result of this restoration 
is   
\begin{eqnarray}
\dis{\int d^4y d^2\t~S\frac{1}{2^7}\tilde{z}_S \Tr~ W_\a(\Vt) W^\a(\Vt)  + 
{\rm H.c.}}\label{GgaugeS}
\end{eqnarray}
Hence, chiral (or antichiral) parts of the vector renormalization 
couplings are of importance only if we say about the renormalization 
of the soft gauge coupling $S.$ This result is in accordance 
with our previous results \cite{AKK} obtained from the analysis of 
divergences in supergraphs. 

The following notation is used for brevity in (\ref{GgaugeS})
\begin{eqnarray*} 
\dis{W^\a\le V\ri \equiv \Db^2 \le e^{-V}D^\a e^V \ri.} 
\end{eqnarray*}   
 
The singular part of the effective action  $\tilde{\G}$ can be written as 
a combination of (\ref{GgaugeS}) and (\ref{depS}),
\begin{eqnarray*}
& \tilde{\G}_{\rm{sing}} = \dis{\int d^4y d^2\t~S\frac{1}{2^7}
~\tilde{z}_S~
\Tr~ W_\a(\Vt) W^\a(\Vt)  + {\rm H.c.}}  \\
& + \dis{\int  d^4x d^2\t d^2\bt~\Tr\frac{1}{32}\frac{\tilde{z}_1\Vt}
{\sqrt{\tilde{\a}}}\le D^2\Db^2 +
 \Db^2D^2\ri \frac{\tilde{z}_1\Vt}{\sqrt{\tilde{\a}}}} \\
& + \dis{\int  d^4x d^2\t d^2\bt~2i~\Tr\le (b + \bar{b})
\frac{\tilde{z}_1}{\sqrt{\tilde{\a}}}
 + \Kt\ri~\left[\del_{\bar{c},c}\Vt\right].}
\end{eqnarray*}

Now we should go back to the initial variables $V$ and $K$, that is, we should 
made the change of variables in $\tilde{\G}$ reversed to (\ref{changeS}).
Hence, the singular part of the effective action which corresponds 
to the theory with the classical action (\ref{SSgauge}) is  
\begin{eqnarray}
& \G_{\rm{sing}} = \dis{\int d^4y d^2\t~S\frac{1}{2^7}~
\tilde{z}_S 
\Tr~ W_\a\le\frac{V}{\tilde{z}_1}\ri W^\a\le\frac{V}{\tilde{z}_1}\ri  + 
{\rm H.c.}}  \no\\
& + \dis{\int  d^4x d^2\t d^2\bt~\Tr\frac{1}{32}\frac{V}
{\sqrt{\tilde{\a}}} \le D^2\Db^2 + \Db^2D^2\ri 
\frac{V}{\sqrt{\tilde{\a}}}}  \label{GsingS}\\
& + \dis{\int  d^4x d^2\t d^2\bt~2i~\Tr\le 
(b + \bar{b})\frac{\tilde{z}_1}{\sqrt{\tilde{\a}}} + K\tilde{z}_1\ri~\left[
\del_{\bar{c},c}\le\frac{V}{\tilde{z}_1}\ri\right].} \no
\end{eqnarray}   
Hence, all divergences can be removed from  $\G_{\rm{sing}}$
by the following rescaling of fields and couplings in the path 
integral (\ref{pathS})
\begin{eqnarray}
V = V_R~\tilde{z}_1, ~~~ S  = S_R~\tilde{z}_S^{-1}  ~~~
\sqrt{\tilde{\a}} = \tilde{z}_1~\sqrt{\tilde{\a}_R}, ~~~  K = K_R~\tilde{z}_1^{-1}.
\end{eqnarray}

\section{Conclusions and Discussions}

In this paper the relations (\ref{rule}) and (\ref{rule2}) between the 
renormalization constants of the softly broken SYM and their prototypes 
from the corresponding rigid theory 
which have been found in Ref. \cite{Jones} starting from the Hisano--Shifman 
nonperturbative result \cite{Shifman} and in Ref. \cite{AKK} starting 
from the supergraph technique for vector vertices have been derived from the 
Slavnov--Taylor identities. It has been shown that  
the modification (\ref{mod}) of the gauge fixing condition is    
necessary and important for the renormalization procedure 
in the softly broken SYM. 

It is clear from the analysis performed here that instead of 
a space-time independent soft gauge coupling we could consider 
any chiral superfield without changing the proof given 
in this paper. This can be important for the models in which 
supersymmetry breaking is communicated to the observable world 
through the interactions with messengers. In these models
$S$ is a messenger superfield which can gain vacuum expectation value 
for its highest component due to interactions with a hidden sector. 
\cite{Giudice,Giudice-2,GMB}. This idea with a toy model for a hidden 
sector has been considered in Ref. \cite{Slavnov}.

As to the relation between chiral matter renormalization constants
of the soft theory and those of the rigid theory, it has been established 
in Ref. \cite{Yam} as substitutions of background superfields into rigid 
renormalization constants instead of rigid couplings.      
The result of these substitutions can be  described as in the Refs.
\cite{Jones,AKK} through differential operators that act in the coupling 
constants space of the rigid theory. The same operators can be used to 
relate soft and rigid renormalization group functions \cite{Jones,AKK}. 
Possible applications of the relations between soft and rigid RG functions 
to the analysis of phenomenological models can be found in Refs. 
\cite{AKK,Giudice,Giudice-2,appl}.
\vskip 3mm 
\noindent {\large{\bf{Acknowledgements}}} 
\vskip 3mm
I am grateful to Antonio Masiero for many discussions. This 
work is supported by INFN.

\section*{Appendix.} 
\setcounter{equation}{0}
\def\theequation{A.\arabic{equation}}
Our supersymmetric notation are 
\begin{eqnarray*}
& \dis{\le\psi \sm \bchi\ri \equiv \psi_\a \sm^{\a\db} \bchi_\db, ~~~ 
\le\psi \sm \bchi\ri^\dg = \le \chi \sm \bpsi\ri,} \\
& \dis{\sm^{\a\db} = \le {\rm I}, ~ \sigma_i \ri, ~~~ \bsm^{~~\db\a} = \sm^{\a\db},} \\
&  \dis{\chi^\a = \e^{\a\b}\chi_\b, ~~~ \e^{12} = -1,} \\
& \dis{ \t^2 = - \t_\a\t^\a,~~~\bt^2 = - \bt^\da\bt_\da ~~~
        \Rar~~~ {\t^2}^\dg = \bt^2,} \\  
& \dis{ \t_\a\t_\b = \f12\e_{\a\b}\t^2, ~~~\Rar~~~ \t^\a\t^\b = -\f12\e^{\a\b}\t^2,} \\
& \dis{ \bt_\da\bt_\db = - \f12\e_{\da\db}\bt^2, ~~~\Rar~~~\bt^\da\bt^\db =  
   \f12\e^{\da\db}\bt^2,} \\
& \dis{\pd^\a\t_\b = \delta^\a_\b~~~\Rar~~~\bt_\db\ol{\pdb^\da} = \delta^\da_\db,} \\
& \dis{\int d^2\t \t^2 \equiv \frac{1}{4}\pd^2 \t^2 = 
  - \pd_\a\pd^\a \t^2 = -1, ~~~  
  \int d^2\bt \bt^2 \equiv \frac{1}{4} \ol{\pdb^2} \bt^2 = 
  - \ol{\pdb^\da}\ol{\pdb_\da}\bt^2  = -1,} \\
& \dis{\le \sm\bsn - \sn\bsm \ri \equiv \smn,} \\ 
& \dis{\sm^{\a\db}(\bsn)_{\db\g} = \eta_{mn}\delta^\a_\g 
+ \f12{\smn^\a}_\g,} \label{Fsmn}\\ 
& \dis{\Tr \le \sm\bsn\sk\bsl \ri = 2 \le \eta_{mn}\eta_{kl} - \eta_{nl}\eta_{mk} 
+ \eta_{ml}\eta_{nk} + i\e_{mnkl} \ri,}  \\
& \dis{\e_{0123} = 1.}
\end{eqnarray*}
The algebra of supersymmetry and covariant derivatives is 
\begin{eqnarray}
& \dis{\ve_\a Q^\a + \Qb^\da\bve_\da = \ve_\a\le\pd^\a + i\sm^{\a\db}\bt_\db\pd_m\ri +
 \le\ol{\pdb^\da} - i\t_\b\sm^{\b\da}\pd_m\ri\bve_\da,} \no\\
&  \dis{Q^\a = \pd^\a + i\sm^{\a\db}\bt_\db\pd_m, ~~~ 
 \Qb^\da = \ol{\pdb^\da} - i\t_\b\sm^{\b\da}\pd_m,} \no\\ 
& \dis{\{Q^\a,\Qb^{\db}\} = -2i\sm^{\a\db}\pd_m,~~~\{Q^\a,Q^{\b}\} = 
  \{\Qb^\da,\Qb^{\db}\} = 0,} \no\\
& \dis{\{D^\a,\Qb^{\db}\} = 0,} \no\\
& \dis{D^\a = \pd^\da - i\le\sm\bt\ri^\a\pd_m, ~~~ 
  \Db^\da = \ol{\pdb^\da} + i\le\t\sm\ri^\da\pd_m,} \no\\
& \dis{\{D^\a,\Db^{\db}\} = 2i\sm^{\a\db}\pd_m,~~~\{D^\a,D^{\b}\} = 
  \{\Db^\da,\Db^{\db}\} = 0,} \no\\ 
&  \dis{\le D^\a\Db^2D_\a \ri^\dg =  D^\a\Db^2D_\a,} \no\\
& \dis{\frac{D^\a\Db^2D_\a}{8\Box} - \frac{D^2\Db^2 + \Db^2D^2}{16\Box} = 1}, 
       \label{proj}\\
& \dis{\Box = \eta_{mn}\pd_m\pd_n = \frac{\pd}{\pd x^0}\frac{\pd}{\pd x^0}   
   - \frac{\pd}{\pd x^1}\frac{\pd}{\pd x^1} - \dots,} ~~~
\dis{\eta_{mn} = \le 1, -1, -1, -1 \ri}. \no
\end{eqnarray}
The classical rigid action $S^{\rm{R}}$  of the supersymmetric theory 
with $N =1$ supersymmetry without soft terms in the 
superfield formalism is 
\begin{eqnarray}
& \dis{\int d^4y d^2\t~\fra1g2\frac{1}{2^7}\Tr~ W_\a W^\a  +  
\int d^4\yb d^2\bt~\fra1g2\frac{1}{2^7}\Tr~ \bW^\da \bW_\da } \nonumber\\
& + \dis{\int d^4x d^2\t d^2\bt ~\bF^i{(e^{V})_i}^j\F_j} + \label{SR}\\
& + \dis{\int d^4yd^2\t \left[y^{ijk}\F_i\F_j\F_k + M^{ij}\F_i\F_j \right]} 
+ \dis{\int d^4\yb d^2\bt \left[\yb_{ijk}\bF^i\bF^j\bF^k \no   
+ \bar{M}_{ij}\bF^i\bF^j \right]}.
\end{eqnarray}
Here $W_\a$ is the supertensity,
\begin{eqnarray*} 
\dis{W^\a \equiv \Db^2 \le e^{-V}D^\a e^V \ri.} 
\end{eqnarray*}
For the real superfield $V$ in the WZ gauge, 
\begin{eqnarray*}
V =  2\t\sm\bt A_m + \t^2 \bla^\da\bt_\da + \bt^2\t_\a\la^\a 
+ \t^2\bt^2D,
\end{eqnarray*}
we have the following results 
\begin{eqnarray*}
& \dis{W^\a  = - 4\le \la^\a - 2\t^\a D + i\f12\t^\b{\smn^\a}_\b F_{mn} 
- i\t^2\Dc_m(\sm\bla)^\a\ri}, \\
& \dis{\int d^4yd^2\t ~\Tr~ W_\a W^\a = \int d^4x \Tr~4^2\le 4D^2 -2F_{mn}F_{mn} 
+ iF_{mn}\tilde{F}_{mn} + 2i\la\sm\Dc_m\bla \ri},
\end{eqnarray*}
where the following notation is used: 
\begin{eqnarray*}
& \dis{F_{mn} \equiv  \pd_mA_n - \pd_nA_m + i[A_m,A_n],} \\  
& \dis{\Dc_m\la^\a \equiv \pd_m\la^\a + i[A_m,\la^\a],} \\
& \dis{\Dc_m\bla^\da \equiv  \pd_m\bla^\da - i[\bla^\da,A_m] = 
\pd_m\bla^\da + i[A_m,\bla^\da]},  \\
& \dis{\Rar \le\Dc_m\la^\a\ri^\dg = \Dc_m\bla^\da},   ~~~ 
\dis{\tilde{F}_{mn} \equiv \e_{mnkl}F_{kl}}. 
\end{eqnarray*}
Hence, for the gauge part of (\ref{SR}) we have the 
component action  
\begin{eqnarray*}
\dis{\int d^4x \left[\frac{1}{2g^2}\Tr~\le 2D^2 - F_{mn}^2 
+ i\la\sm\Dc_m\bla \ri \right].}
\end{eqnarray*}
All fields of the real supermultiplet are in the adjoint representation of 
the gauge group
\begin{eqnarray*}
W_\a = W_\a^a T^a,   ~~~  \Tr~ \le T^aT^b \ri = \f12\delta^{ab}, 
~~~  \le T^a \ri^\dg = T^a.
\end{eqnarray*}

To fix the gauge we have to add the gauge fixing term and the ghost terms to the 
action (\ref{SR}) which can be chosen in the standard form \cite{West} 
\begin{eqnarray}
& \dis{\int d^4x d^2\t d^2\bt ~\frac{1}{16}\frac{1}{\a}\Tr~\le\Db^2 V\ri\le D^2 V\ri}
 \label{gfgR}\\ 
& + \dis{\int d^4y d^2\t~\frac{i}{2}\Tr~  b~\Db^2 \del_{\bar{c},c}V  + 
\int d^4\yb d^2\bt~\frac{i}{2}\Tr~  \bar{b}~D^2\del_{\bar{c},c} V.} \no
\end{eqnarray}
where $b$ and $\bar{b}$ are the antighost chiral and antichiral superfields, 
and $c$ and $\bar{c}$ are the ghost chiral and antichiral superfields. In case 
if $\a = 1$ we have Feynman's gauge fixing term. Such a choice of the gauge fixing 
and the ghost terms means that we fix the gauge arbitrariness by imposing the 
condition 
\begin{eqnarray*}
D^2 V(x,\t,\bt) = \bar{f}(\yb,\bt),  ~~~  \Db^2 V(x,\t,\bt) = f(y,\t),  
\end{eqnarray*}
where  $\bar{f}$ and $f$ are arbitrary chiral and antichiral functions.
Under the gauge transformation the vector superfield $V$ transforms 
as
\begin{eqnarray}
\dis{e^V \rar e^{\bLac} e^V e^{\Lac}}, \label{Vgauge}
\end{eqnarray} 
where  $\bLac,\Lac$ are antichiral and chiral degrees of gauge freedom.
We define $\del_{\bLac,\Lac}V$ as the solution to the equation 
\begin{eqnarray*}
\dis{e^{V + \del_{\bLac,\Lac}V} =  e^{\bLac} e^V e^{\Lac}},
\end{eqnarray*}
with infinitesimal fields $\bLac,\Lac.$ This equation can be transformed 
to the form 
\begin{eqnarray}
e^V \le \del_{\bLac,\Lac}V \ri - \le \del_{\bLac,\Lac}V \ri e^V = 
[V,\bLac] e^V + e^V [V,\Lac] \label{delV} 
\end{eqnarray} 
that can be solved \cite{West} as 
\begin{eqnarray*}
\del_{\bLac,\Lac}V = \frac{V}{2}\coth\frac{V}{2}\wedge \le \bLac + \Lac \ri
- \frac{V}{2}\wedge \le \bLac - \Lac \ri.
\end{eqnarray*} 
Hence, the total gauge part of the classical action (\ref{SR}) is  
\begin{eqnarray}
& S^{\rm{R}}_{\rm{gauge}} = \dis{\int d^4y d^2\t~\fra1g2\frac{1}{2^7}\Tr~ W_\a W^\a  +  
\int d^4\yb d^2\bt~\fra1g2\frac{1}{2^7}\Tr~ \bW^\da \bW_\da } \no \\
& + \dis{\int d^4x d^2\t d^2\bt ~\frac{1}{16}\frac{1}{\a}\Tr~\le\Db^2 V\ri\le D^2V\ri}
  \label{SRgauge} \\ 
& + \dis{\int d^4y d^2\t~\frac{i}{2}\Tr~  b~\Db^2 \del_{\bar{c},c}V  + 
\int d^4\yb d^2\bt~\frac{i}{2}\Tr~  \bar{b}~D^2\del_{\bar{c},c} V.} \no
\end{eqnarray}

Below we concentrate on the gauge part of the action. The short review 
of the procedure necessary to remove divergences from the effective action 
is given. This review is necessary to compare with the case of softly broken 
supersymmetry analyzed in the main part of this paper. This review is very concise 
and everybody who is interested in more details can refer to the 
reviews \cite{Piguet,Becchi}. The BRST symmetry is reviewed in \cite{Becchi}
and applications of Slavnov--Taylor identities to the renormalization        
of supersymmetric theories can be found in \cite{Piguet}. 

The action (\ref{SRgauge}) is invariant under the BRST symmetry,
\begin{eqnarray}
& \dis{e^V \rar e^{i\bar{c}\ve} e^V e^{ic\ve}},& ~~~~  
\dis{\del b = \frac{1}{32}\frac{1}{\a}\le\Db^2D^2V\ri\ve} \no \\
& \dis{c \rar c + ic^2\ve},& ~~~~ \dis{\del \bar{b} = \frac{1}{32}\frac{1}{\a}
\le D^2\Db^2V\ri\ve}, \label{BRSTr} \\
& \dis{\bar{c} \rar \bar{c} - i\bar{c}^2\ve}, & ~~~~  \no
\end{eqnarray}
with an Hermitian Grassmannian parameter $\ve$, $\ve^\dg = \ve.$ 
This looks like a gauge transformation for the vector superfield (\ref{Vgauge}). 
The transformation of the ghost superfields is caused by the transformation of
$\del_{\bar{c},c}V$ under the BRST transformation of $V$  in (\ref{BRSTr}).  
By construction, $\del_{\bar{c},c}V$ is the solution to the equation (\ref{delV})
when $\bLac,\Lac$ are replaced with $\bar{c},c$ respectively. If in the equation 
(\ref{delV}) we put the transformed vector superfield $V + \del_{i\bar{c}\ve,ic\ve}V$  
according to 
\begin{eqnarray*}
\dis{e^{V + \del_{i\bar{c}\ve,ic\ve}V} =  e^{i\bar{c}\ve} e^V e^{ic\ve}}
\end{eqnarray*}
instead of $V,$ 
we get that the solution $\del_{\bar{c},c}V$ to eq. (\ref{delV}) takes the 
transformation  $\del\le\del_{\bar{c},c}V\ri$ that satisfies to the the 
equation 
\begin{eqnarray*}
e^V \le \del\le\del_{\bar{c},c}V\ri \ri - \le \del\le\del_{\bar{c},c}V\ri \ri e^V = 
[V,i\bar{c}^2\ve] e^V + e^V [V,-ic^2\ve].  
\end{eqnarray*} 
The transformations of the ghost superfields in (\ref{BRSTr}) compensate
this transformation of the $\del_{\bar{c},c}V,$ so that the total BRST 
transformation of $\del_{\bar{c},c}V$ is vanishing,  
\begin{eqnarray*}
\del_{\rm{BRST}}\le\del_{\bar{c},c}V\ri = 0.
\end{eqnarray*}  
At the same time, the transformation of antighost superfields $b,\bar{b}$ 
is necessary to remove the non-invariance of the gauge fixing term. 

The path integral for the rigid theory is defined as   
\begin{eqnarray}
& \dis{Z[J,\eta,\bar{\eta},\rho,\bar{\rho},K,L,\bar{L}] = 
\int dV~dc~d\bar{c}~db~d\bar{b}~\exp i}\left[\dis{S^{\rm{R}}_{\rm{gauge}}} 
\right.  \label{pathR}\\
& \left. + \dis{2~\Tr\le JV + i\eta c + i\bar{\eta}\bar{c} + i\rho b + 
i\bar{\rho}\bar{b}\ri + 
2~\Tr\le iK\del_{\bar{c},c}V + Lc^2 + \bar{L}\bar{c}^2 \ri}\right]. \no
\end{eqnarray}
The third term in the brackets is the BRST invariant since the external 
superfields $K$ and $L$ are BRST invariant by definition. All fields
in the path integral are in the adjoint representation of the gauge group.
For the sake of brevity we omit the symbol of integration in the 
terms with external sources, keeping in mind that it is the full superspace 
measure for vector superfields and the chiral measure for  
chiral superfields. 

Having made the change of fields in the path integral 
\begin{eqnarray*}
b \rar b + \ve,    ~~~   \bar{b} \rar  \bar{b} + \bve 
\end{eqnarray*}
with an arbitrary chiral superfield $\ve,$  two identities can be 
obtained   
\begin{eqnarray}
\bar{\rho} - i\frac{1}{4}D^2\frac{\del W}{\del K} = 0,  ~~~
\rho - i\frac{1}{4}\Db^2\frac{\del W}{\del K} = 0, \label{ghostE}
\end{eqnarray}
where the standard definition for the connected diagrams generator
is used, 
\begin{eqnarray*}
Z = e^{-iW}.
\end{eqnarray*}
For the derivative with respect to vector superfield we use the 
definition 
\begin{eqnarray*}
\frac{\del}{\del K} \equiv T^a \frac{\del}{\del K^a},
\end{eqnarray*}
while the derivative with respect to chiral superfield is 
defined from the requirement 
\begin{eqnarray*}
\frac{\del}{\del \eta(y,\t)}\int d^4y' d^2\t'~2\Tr~\eta(y',\t')c(y',\t') = c(y,\t) 
\Rar \frac{\del\eta^a(y',\t')}{\del \eta^b(y,\t)} = \frac{1}{4}\Db^2\del^{(8)}(z-z')
\del^{ab}.
\end{eqnarray*}
Here $z$ is the definition for the total superspace coordinate $z = (x,\t,\bt),$
so that 
\begin{eqnarray*}
\del^{(8)}(z-z') =  \del^{(4)}(x-x')~\del^{(2)}(\t-\t')~\del^{(2)}(\bt-\bt').
\end{eqnarray*}
The effective action $\G$ is related to $W$ by the Legendre 
transformation
\begin{eqnarray}
& \dis{V \equiv - \frac{\del W}{\del J}, ~~ ic  \equiv - \frac{\del W}{\del \eta}, ~~
i\bar{c}  \equiv - \frac{\del W}{\del \bar{\eta}}, ~~ 
ib \equiv - \frac{\del W}{\del \rho}, ~~i\bar{b} \equiv - 
\frac{\del W}{\del \bar{\rho}}}, \label{defphi} \\
& \dis{\G = - W - 2~\Tr\le JV + i\eta c + i\bar{\eta}\bar{c} + i\rho b + 
i\bar{\rho}\bar{b}\ri \equiv - W - 2~\Tr\le X\phi\ri}, \label{Legendre} \\
&  \dis{\le X\phi\ri \equiv i^{G(k)}X^{k}\phi^{k}}, \no\\
& \dis{X \equiv \le J,\eta,\bar{\eta},\rho,\bar{\rho}\ri,  ~~~   \phi \equiv 
\le V,c,\bar{c},b,\bar{b} \ri}, \no 
\end{eqnarray}
where $G(k) = 0$ if $\phi^{k}$ is the Bose superfield and $G(k) = 1$
if $\phi^{k}$ is the Fermi superfield. Iteratively all equations (\ref{defphi})
can be reversed, 
\begin{eqnarray*}
X = X[\phi,K,L,\bar{L}],
\end{eqnarray*}
and the effective action is defined in terms of new variables, 
$\G = \G[\phi,K,L,\bar{L}].$ 
Hence, the following equalities take place
\begin{eqnarray}
& \dis{\frac{\del \G}{\del V} = - 
\frac{\del X^{a}}{\del V}\frac{\del W}{\del X^{a}} -
i^{G(a)}\frac{\del X^{a}}{\del V}\phi^{a} - J = - J,} \no\\
& \dis{\frac{\del \G}{\del K} = 
- \frac{\del X^{a}}{\del K}\frac{\del W}{\del X^{a}} - 
i^{G(a)}\frac{\del X^{a}}{\del K}\phi^{a} - \frac{\del W}{\del K} = 
- \frac{\del W}{\del K},} \label{GW}\\
& \dis{\frac{\del \G}{\del c} = i\eta, ~~ \frac{\del \G}{\del\bar{c}} = i\bar{\eta}, ~~
\frac{\del \G}{\del b} = i\rho, ~~ \frac{\del \G}{\del\bar{b}} = i\bar{\rho},  ~~
\frac{\del \G}{\del L} = - \frac{\del W}{\del L}, ~~  
\frac{\del \G}{\del \bar{L}} = - \frac{\del W}{\del \bar{L}}}. \no
\end{eqnarray}
Here all Grassmannian derivatives are left derivatives. Therefore, the ghost equations 
(\ref{ghostE}) can be written as     
\begin{eqnarray}
\frac{\del \G}{\del\bar{b}}  - \frac{1}{4}D^2\frac{\del \G}{\del K} = 0,  ~~~
\frac{\del \G}{\del b} - \frac{1}{4}\Db^2\frac{\del \G}{\del K} = 0. \label{ghostEr}
\end{eqnarray}
If the change of fields (\ref{BRSTr}) in the path integral (\ref{pathR}) is made,
that we get the Slavnov--Taylor identity as the result of invariance of the 
integral (\ref{pathR}) under a change of variables,
\begin{eqnarray*}
& \Tr\left[\dis{J\frac{\del}{\del K} - i\eta\le\frac{1}{i}\frac{\del}{\del L}\ri + 
i\bar{\eta}\le\frac{1}{i}\frac{\del}{\del \bar{L}}\ri + 
i\rho\le\frac{1}{32}\frac{1}{\a}\Db^2D^2\frac{\del}{\del J}\ri} \right. \\
& \left. + 
\dis{i\bar{\rho}\le\frac{1}{32}\frac{1}{\a}D^2\Db^2\frac{\del}{\del J}\ri}\right]W =0,
\end{eqnarray*}
or, taking into account the relations (\ref{GW}), we have 
\begin{eqnarray}
& \Tr\left[\dis{\frac{\del \G}{\del V}\frac{\del \G}{\del K} 
 - i \frac{\del \G}{\del c}\frac{\del \G}{\del L}  
+ i \frac{\del \G}{\del \bar{c}}\frac{\del \G}{\del \bar{L}} 
- \frac{\del \G}{\del b}\le\frac{1}{32}\frac{1}{\a}\Db^2D^2V\ri} \right. \label{STr}\\ 
& \left. - \dis{\frac{\del \G}{\del \bar{b}}\le\frac{1}{32}\frac{1}{\a}D^2\Db^2V\ri}
\right]=0. \no
\end{eqnarray}
The identities (\ref{ghostEr}) and (\ref{STr}) allow us to remove all possible
divergences from the effective action $\G$ by rescaling superfields and 
couplings in the classical action (\ref{SRgauge}). Indeed, the identity
(\ref{ghostEr}) restricts the dependence of $\G$ on the antighost 
superfields and on the external source $K$ to an arbitrary dependence on their 
combination $b + \bar{b} + K.$ This means that the corresponding 
singular part of the effective action is 
\begin{eqnarray*}
\int  d^4x d^2\t d^2\bt~2i~\Tr\le b + \bar{b} + K\ri~A(x,\t,\bt),
\end{eqnarray*}
where  $A(x,\t,\bt)$ is a combination of $c,\bar{c},V.$ By index counting arguments
we know that the singular part repeats the structure of the classical action 
(\ref{SRgauge}) up to coefficients. Hence, $A(x,\t,\bt)$ starts from the 
$z_1(c + \bar{c}),$ since $\G$ is Hermitian. Here $z_1$ is a  constant that 
can be found by using the supergraph technique. The Slavnov--Taylor identity
(\ref{STr}) fixes the coefficient before the longitudinal part of the 2-point vector 
Green's function. Indeed, by using projectors from (\ref{proj}) the 2-point vector 
correlator can be decomposed as  
\begin{eqnarray}
& \dis{V\le D,\Db,D,\Db \ri V = V\le D,\Db,D,\Db \ri \frac{D^\a\Db^2D_\a}{8\Box} V} 
\label{decR} \\  
& - \dis{V \le D,\Db,D,\Db \ri\frac{D^2\Db^2 + \Db^2D^2}{16\Box}V}, \no
\end{eqnarray}     
where the four derivatives in parenthesis can stand in some (in general, unknown) 
way. The identity
(\ref{STr}) means that these four derivatives in the second term of this decomposition
must cancel the $\Box$ in the denominator, and the second term is reduced to 
the form  
\begin{eqnarray*}
z_2 \frac{1}{\a}\frac{1}{32}V\le D^2\Db^2 + \Db^2D^2\ri V.   
\end{eqnarray*}
The Slavnov--Taylor identity also gives that $z_2 =1,$ that is there is no 
infinite correction to the longitudinal part of the 2-point vector function.
The same arguments can be applied even in the case of the total effective 
action, taking into account the whole dependence of the effective action 
$\G$ on the combination $b +\bar{b} +K.$ Hence, there is no finite correction to 
the longitudinal part of the 2-point vector correlator. 

Now it is necessary to consider contributions into $A(x,\t,\bt)$ of the next orders 
in fields. For example, the third order terms can be presented as 
\begin{eqnarray}
& \dis{\int  d^4x d^2\t d^2\bt~2i~\Tr\le b + \bar{b} + K\ri~\left[
z_1(c + \bar{c}) + z_4 \le Vc + \bar{c}V \ri + z_5\le cV + V\bar{c}\ri \right]} 
\label{3odR}\\
& + \dis{\int d^4y d^2\t~2\Tr~z_6Lc^2 + 
\int d^4\yb d^2\bt~2\Tr~\bar{z}_6\bar{L}\bar{c}^2} \no
\end{eqnarray}
By the no-renormalization theorem for the superpotential \cite{West} we get 
\begin{eqnarray*}
z_6 = \bar{z}_6 = 1.
\end{eqnarray*}
To fix the constants $z_4$ and $z_5$, we make the change of variables
in the effective action $\G,$
\begin{eqnarray}
& \dis{\G\left[V,c,\bar{c},b,\bar{b},K,L,\bar{L}\right] = 
\G\left[V(\Vt),c,\bar{c},b,\bar{b},K(\Kt),L,\bar{L}\right] = 
\tilde{\G}\left[\Vt,c,\bar{c},b,\bar{b},\Kt,L,\bar{L}\right]}, \no\\
& \dis{V = \Vt z_1,  ~~~  K = \frac{\Kt}{z_1}.} \label{changeR}
\end{eqnarray}
The Slavnov--Taylor identity (\ref{STr}) in new variables is 
\begin{eqnarray}
& \Tr\left[\dis{\frac{\del \tilde{\G}}{\del \Vt}\frac{\del \tilde{\G}}{\del \Kt} 
 - i \frac{\del \tilde{\G}}{\del c}\frac{\del \tilde{\G}}{\del L}  
+ i \frac{\del \tilde{\G}}{\del \bar{c}}\frac{\del\tilde{\G}}{\del \bar{L}} 
- \frac{\del\tilde{\G}}{\del b}\le\frac{1}{32}\frac{1}{\a}\Db^2D^2\Vt z_1\ri} \right. 
\label{STrm}\\ 
& \left. - \dis{\frac{\del \tilde{\G}}{\del \bar{b}}
\le\frac{1}{32}\frac{1}{\a}D^2\Db^2\Vt z_1\ri}
\right]=0. \no
\end{eqnarray}
The part of the effective action (\ref{3odR}) in the new variables looks like 
\begin{eqnarray}
& \dis{\int  d^4x d^2\t d^2\bt~2i~\Tr\le (b + \bar{b})z_1 + \Kt\ri~\left[
(c + \bar{c}) + z_4' \le \Vt c + \bar{c}\Vt \ri + z_5'\le c\Vt + \Vt\bar{c}\ri \right]} 
\no  \\
& + \dis{\int d^4y d^2\t~2\Tr~Lc^2 + 
\int d^4\yb d^2\bt~2\Tr~\bar{L}\bar{c}^2}, \label{3odRm}
\end{eqnarray}
where $z_4'$ and $z_5'$ are new constants. 

The higher order terms in the brackets of (\ref{3odRm}) are restored 
unambiguously by themselves in the iterative way due to the first three terms 
in the modified identities (\ref{STrm}). As the result we have 
\begin{eqnarray}
\dis{\int  d^4x d^2\t d^2\bt~2i~\Tr\le (b + \bar{b})z_1 + \Kt\ri~\left[
\del_{\bar{c},c}\Vt\right].}  \label{depR}
\end{eqnarray}

Now it is necessary to consider the transversal part of the 2-point 
vector correlator. Having made the change of variables (\ref{changeR}) 
in the effective action, we get the first term in the decomposition   
(\ref{decR}) as 
\begin{eqnarray}
\int d^4x d^2\t d^2\bt z_3~z_1^2~\fra1g2\frac{1}{2^5}
\Tr~D_\a \Vt \Db^2D^\a \Vt  + {\rm H.c.}.  \label{2pointR} 
\end{eqnarray}
This is the only gauge invariant combination fixed by the first term 
in the modified identities (\ref{STrm}), if we take into account 
already fixed dependence (\ref{depR}) of the singular part of $\tilde{\G}$ on the
external source $\Kt.$ It means that the four derivatives into the 
first term of the decomposition (\ref{decR}) cancel the D'Alambertian 
in the denominator. Here $z_3$ is a  constant that can be found by using 
the supergraph technique \cite{GRS}. 

The first term in the modified identity (\ref{STrm}) will restore in the 
iterative way higher order terms starting from the bilinear transversal 
2-point correlator (\ref{2pointR}). Hence, the result of this restoration 
is   
\begin{eqnarray}
\dis{\int d^4y d^2\t~\fra1g2\frac{1}{2^7}z_3~z_1^2~\Tr~ W_\a(\Vt) W^\a(\Vt)  +
{\rm H.c.}} \label{GgaugeR}
\end{eqnarray}

The singular part of the effective action  $\tilde{\G}$ can be written as 
a  combination of (\ref{GgaugeR}) and (\ref{depR}),
\begin{eqnarray*}
& \tilde{\G}_{\rm{sing}} = \dis{\int d^4y d^2\t~\fra1g2\frac{1}{2^7}z_3~z_1^2~
\Tr~ W_\a(\Vt) W^\a(\Vt)  + {\rm H.c.}}  \\
& + \dis{\int  d^4x d^2\t d^2\bt~\Tr\frac{1}{\a}\frac{1}{32}z_1\Vt\le D^2\Db^2 +
 \Db^2D^2\ri z_1\Vt} \\
& + \dis{\int  d^4x d^2\t d^2\bt~2i~\Tr\le (b + \bar{b})z_1 + \Kt\ri~\left[
\del_{\bar{c},c}\Vt\right].}
\end{eqnarray*}

Now we should go back to the initial variables $V$ and $K,$ that is,
we should  made the change of variables in $\tilde{\G}$ reversed to 
(\ref{changeR}). Hence, the singular part of the effective action which corresponds 
to the theory with the classical action (\ref{SRgauge}) is  
\begin{eqnarray}
& \G_{\rm{sing}} = \dis{\int d^4y d^2\t~\fra1g2\frac{1}{2^7}z_3~z_1^2~
\Tr~ W_\a\le\frac{V}{z_1}\ri W^\a\le\frac{V}{z_1}\ri  + {\rm H.c.}}  \no\\
& + \dis{\int  d^4x d^2\t d^2\bt~\Tr\frac{1}{\a}\frac{1}{32}V\le D^2\Db^2 +
 \Db^2D^2\ri V} \label{GsingR}\\
& + \dis{\int  d^4x d^2\t d^2\bt~2i~\Tr\le (b + \bar{b})z_1 + K z_1\ri~\left[
\del_{\bar{c},c}\le\frac{V}{z_1}\ri\right].} \no
\end{eqnarray}   
Hence, all possible divergences can be removed from the $\G_{\rm{sing}}$
by the following rescaling of fields and couplings in the path 
integral (\ref{pathR})
\begin{eqnarray*}
V = V_R~z_1, ~~~ \frac{1}{g^2} = \frac{1}{g_R^2}z_1^{-2}z_3^{-1}, ~~~
\a = z_1^2~\a_R,~~~  b = b_R~z_1^{-1},~~~  K = K_R~z_1^{-1}.
\end{eqnarray*}

\end{document}